%% file: main.tex
\documentclass[conference]{IEEEtran}
\IEEEoverridecommandlockouts
\usepackage{cite}
\usepackage{amsmath,amssymb,amsfonts}
\usepackage{graphicx}
\usepackage{textcomp}
\usepackage{xcolor}
\def\BibTeX{{\rm B\kern-.05em{\sc i\kern-.025em b}\kern-.08em
    T\kern-.1667em\lower.7ex\hbox{E}\kern-.125emX}}

\usepackage{subcaption}
\usepackage{algorithm}
\usepackage{algpseudocode}
\usepackage{enumitem}
\usepackage{multirow}
\usepackage{tabularx}
\usepackage{threeparttable}
\usepackage{xspace}
\usepackage{balance}
\usepackage{booktabs}
\usepackage{url}

\begin{document}

\title{MuSeR: Scalable Long-sequence Recommendation with Multi-interest Modeling%
\thanks{* Corresponding author: Maolin Wang (morin.wang@my.cityu.edu.hk).}
}

\author{%
\IEEEauthorblockN{%
Yongkang Fu\textsuperscript{1},
Beining Bao\textsuperscript{2},
Yu Jiang\textsuperscript{3},
Xiangyu Zhao\textsuperscript{2},
Hongyang Wei\textsuperscript{1},
Guangxing Chen\textsuperscript{1},
Zuodong Yang\textsuperscript{1},\\
Shantao Li\textsuperscript{1},
Zonggang Wu\textsuperscript{1},
Yuqi Lu\textsuperscript{1},
Shouke Qin\textsuperscript{1},
Hanmeng Liu\textsuperscript{1},
Maolin Wang\textsuperscript{2,*}}
\IEEEauthorblockA{\textsuperscript{1}\textit{Baidu}, Beijing, China}
\IEEEauthorblockA{\textsuperscript{2}\textit{City University of Hong Kong}, Hong Kong SAR, China}
\IEEEauthorblockA{\textsuperscript{3}\textit{Chinese University of Hong Kong}, Hong Kong SAR, China}
}
\maketitle

\begin{abstract}
Ultra-long user behavior sequences carry rich signals of stable and diverse
preferences, yet industrial recommender systems typically truncate histories to
a few hundred actions under strict latency and memory budgets, leaving
long-term interests under-utilized. Users also pursue multiple heterogeneous
intents across modalities such as news, Q\&A, and short video, which sparse ID
embeddings alone struggle to represent. We present \textbf{Mu}lti-interest
\textbf{Se}quence \textbf{R}epresentation (MuSeR), a retrieval framework built
on the deployed MGS system, which integrates three components: (i)
\emph{hierarchical temporal compression}, which retains recent actions at full
resolution while progressively pooling older segments, so that per-user
histories of $10^{4}$--$10^{5}$ interactions fit within a fixed serving budget;
(ii) \emph{disentangled multi-query interest extraction} with orthogonality
regularization; and (iii) \emph{multimodal semantic alignment}, which augments
sparse item IDs with textual summaries distilled from a large language model.
For industrial deployment, MuSeR further adopts asynchronous
user-representation refresh with adaptive caching and hierarchical beam-search
retrieval across heterogeneous hardware. On three public benchmarks and a
large-scale industrial dataset, MuSeR consistently improves Recall@$K$ over
strong long-sequence and multi-interest baselines. In online A/B tests on Baidu
APP's homepage feed, discovery feed, and short-video scenarios, MuSeR yields
$+0.26\%$ daily active users and $+0.89\%$ total session duration (both
statistically significant, $p<0.05$), alongside reduced serving latency and
cost. Rather than proposing a new modeling primitive, our contribution is a
system-level integration that makes long-term, multi-interest, and multimodal
modeling jointly deployable in a real-time production pipeline, together with
the engineering practices required to sustain it.
\end{abstract}

\begin{IEEEkeywords}
User Interest Modeling, Ultra-long Behavioral Sequences, Recommender Systems, Multi-interest Modeling, Industrial Recommendation
\end{IEEEkeywords}

\input{Contents/INTRODUCTION}
\input{Contents/METHODOLOGY}

\input{Contents/Deployment}
\input{Contents/EXPERIMENT}
\input{Contents/RELATEDWORKS}
\input{Contents/CONCLUSION}

\balance
\bibliographystyle{IEEEtran}
\bibliography{sample-base}

\appendices

\input{Contents/APPENDIX}
\end{document}

%% file: Contents/INTRODUCTION.tex
\section{Introduction}
In recent years, large-scale personalized recommendations have become a cornerstone of mobile super-app ecosystems. In the Baidu APP, the first recommendation homepage, discovery feed, and short-video scenes collectively serve hundreds of millions of daily active users, accounting for the majority of total session duration. Delivering high-quality recommendations in such complex environments requires the system to process user intents with millisecond-level responsiveness while sustaining ultra-high throughput across heterogeneous hardware infrastructures~\cite{covington2016deep}. At the core of personalization is the accurate modeling of user behavioral sequences. Over years of interactions, users accumulate up to $10^{5}$ actions across modalities such as news, QA, and short-video streams, forming ultra-long behavioral logs that encode rich, nuanced, and stable preferences. However, due to online latency constraints and resource limitations, current production systems typically truncate these to only a few hundred recent actions, leading to under-utilization of long-term interests and a reduced ability to capture stable engagement signals and diversified preferences~\cite{chai2025longer, pan2024survey}.

Despite progress in deep sequence modeling and retrieval~\cite{pan2024survey}, several unresolved challenges still hinder industrial-scale deployment of ultra-long sequence modeling. First, computational scalability and efficiency remain major bottlenecks: transformer-based models scale quadratically with input length, making histories of $10^{5}$ actions per user infeasible under typical production SLAs (usually $<$\,300\,ms)~\cite{chai2025longer}. Second, interest diversity and temporal drift amplify modeling complexity: users often exhibit multiple, sometimes conflicting interests across heterogeneous modalities, where long-term preferences evolve slowly but short-term intentions shift rapidly, requiring a unified mechanism to capture both stability and novelty. Third, an alignment gap between sparse IDs and multimodal content limits representation quality: most approaches rely on sparse item embeddings and underutilize semantic signals from text or video, so sequence compression can lose critical semantics, particularly in cold-start and short-video scenarios. Finally, engineering and deployment constraints further complicate adoption: practical recommender systems must balance accuracy gains with asynchronous computation, caching, and load-adaptive serving to ensure cost-efficient, reliable operation across heterogeneous hardware.

Recognizing these challenges, we present \textbf{Mu}lti-interest \textbf{Se}quence \textbf{R}epresentation (MuSeR), an extension of the MGS (Multi-target Graph-based Search) retrieval framework that unifies \textbf{ultra-long sequence compression}, \textbf{multi-query disentangled interest modeling}, and \textbf{multimodal semantic alignment}. MuSeR advances industrial recommendation along three principal dimensions.

First, we design a \textbf{hierarchical temporal compression mechanism} that condenses ultra-long user histories into compact yet informative sequences. Recent actions are preserved with full granularity, mid-term behaviors are downsampled with moderate pooling, while early histories are aggressively compressed into coarser representations. This strategy ensures that the system remains scalable while still covering both short-term and long-term behavioral patterns~\cite{xu2025multi, chai2025longer}.
Second, we introduce a \textbf{multi-query interest extractor}, inspired by query-based representation learning frameworks. Instead of producing a single aggregated embedding, MuSeR learns multiple query vectors that distill different aspects of a user's preferences into several distinct embeddings~\cite{lian2021multi, zhang2020multi, cheng2024accurate}. During training, each embedding is responsible for predicting diverse future actions; in serving, the framework dynamically selects the most relevant interest vector to interact with retrieval candidates. This mechanism allows the system to capture heterogeneous and even conflicting user intents across homepage browsing, discovery feeds, and short-video scenarios~\cite{wu2024when, meng2025user}.
Third, MuSeR integrates \textbf{multimodal alignment} to bridge the gap between sparse item identifiers and rich content semantics. Beyond ID-level embeddings, we incorporate textual summaries generated by ERNIE-4.0-Turbo (distilled into its lightweight, cost-efficient variant) and semantic representations learned through BGE embeddings. The result is a unified multimodal representation that enhances recall quality, strengthens cold-start handling, and improves recommendations in multi-domain environments.

\begin{figure}[t!]
    \centering
    \begin{subfigure}[b]{0.23\textwidth}
        \centering
        \includegraphics[width=0.6\textwidth]{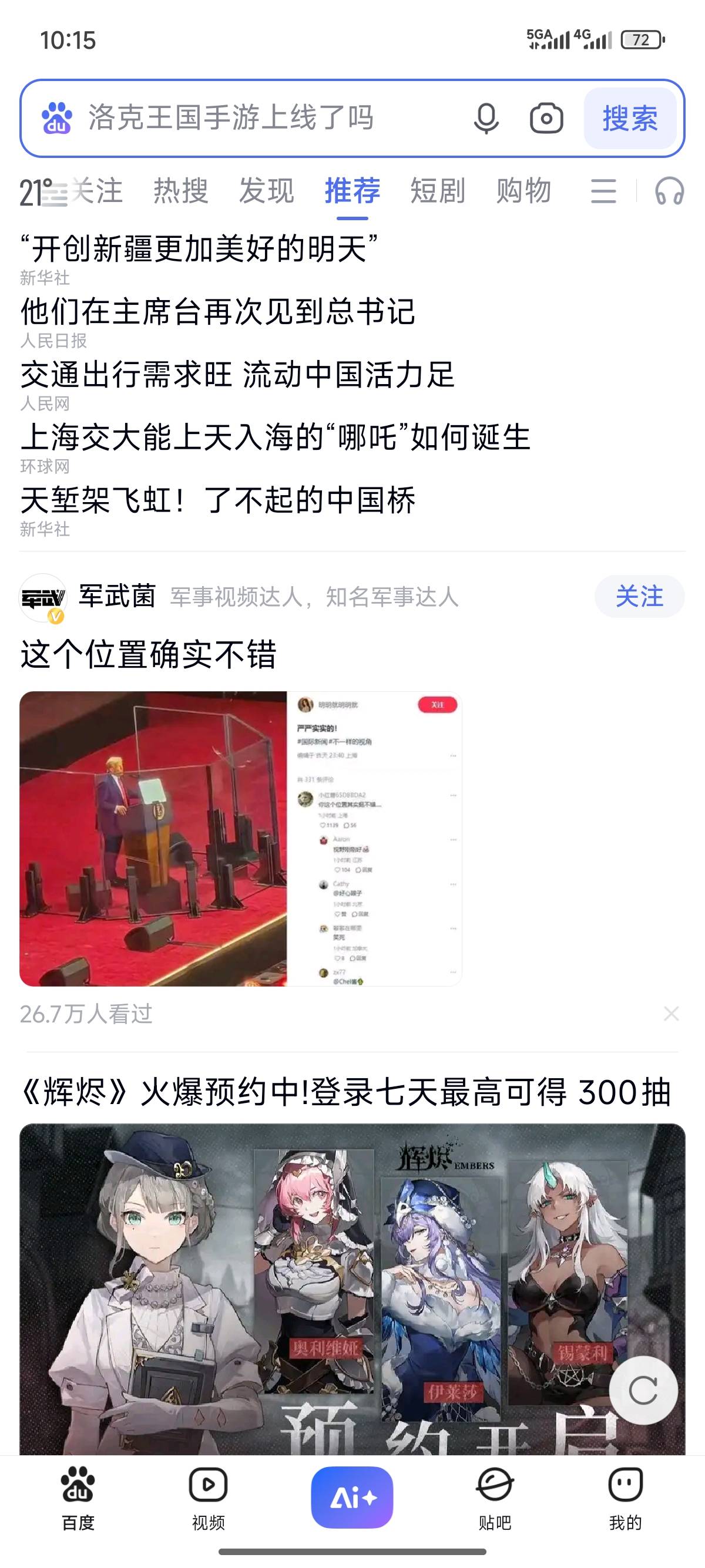}
        \caption{Homepage "Recommended"}
        \label{fig:scene_home}
    \end{subfigure}
    \hfill
    \begin{subfigure}[b]{0.23\textwidth}
        \centering
        \includegraphics[width=0.6\textwidth]{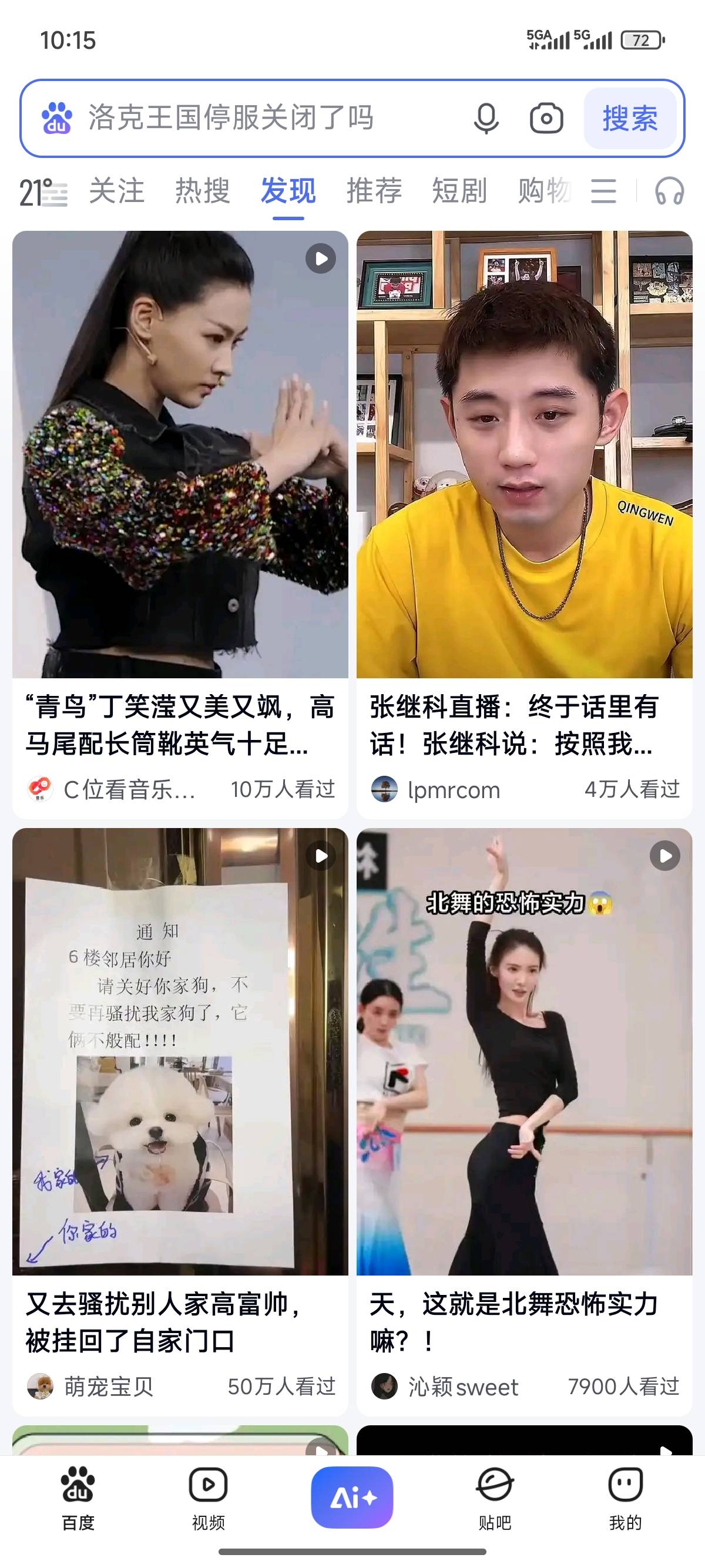}
        \caption{Discovery and short-video}
        \label{fig:scene_short}
    \end{subfigure}
\caption{Deployment scenarios of MuSeR in the Baidu APP and its position in the recommendation pipeline. (a) The homepage ``Recommended'' feed, where MuSeR serves as the retrieval module to recall high-quality candidates from ultra-long behavior logs. (b) The Discovery feed and short-video entrance, where MuSeR supports multi-interest matching under heterogeneous modalities and shifting intents.}
    \label{fig:scenarios}
\end{figure}

From a system engineering perspective, MuSeR couples model innovation with a cost-aware deployment pipeline. Through an \emph{asynchronous computation strategy}, ultra-long sequence representations are periodically refreshed offline or via cache, with refresh frequencies dynamically adjusted by QPS and service load, while online scoring only processes short sequences. In retrieval, we adopt a \emph{hierarchical beam-search strategy} adapted from HNSW, combining efficient vector similarity with deep multi-target scoring. This hybrid design ensures high recall without compromising latency, and is well-suited for heterogeneous CPU/GPU clusters in real production environments.

\begin{figure*}[t]
    \centering
    \includegraphics[width=\textwidth]{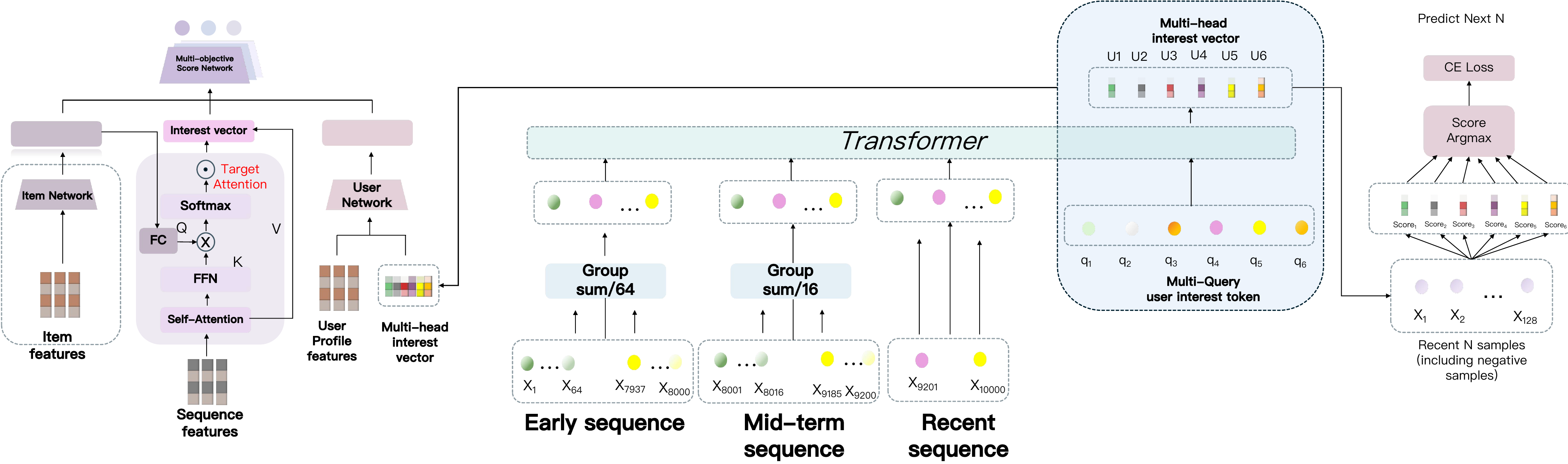}
    \caption{
        Overview of the proposed \textbf{MuSeR} framework, 
        which integrates hierarchical temporal compression, multi-query interest extraction, 
        and multimodal semantic alignment for efficient user modeling and retrieval.
    }
    \label{fig:architecture}
\end{figure*}

We validated MuSeR through extensive A/B testing in Baidu APP's homepage, discovery feed, and short-video scenarios~(as shown in Figure~\ref{fig:scenarios}). \textbf{Since August 2025, MuSeR has been fully deployed in production}, demonstrating consistent online improvements with feed DAU lifted by $+$0.26\% and total session duration increased by $+$0.89\%. Beyond numerical gains, MuSeR effectively leveraged per-user histories of up to $10^{5}$ interactions to surface deeper, more stable interests, and successfully applied multimodal alignment to improve semantic recall, confirming its scalability and robustness at industry scale.
Specifically, we make the following contributions:

\begin{itemize}[leftmargin=*]
\item We propose \textbf{MuSeR}, a novel end-to-end long-sequence recommendation framework that unifies hierarchical compression, multi-query interest extraction, and multimodal semantic alignment~\cite{yan2024trinity}.
\item We design a hierarchical temporal compression mechanism and a multi-query extractor that jointly capture both long-term stable interests and short-term evolving preferences, improving user modeling depth~\cite{wang2023incremental, li2025denoising}.
\item We develop a robust multimodal alignment pipeline based on ERNIE-4.0-Turbo and BGE, mapping sparse item IDs into dense semantic vectors that enhance retrieval quality, particularly in cold-start and short-video scenarios.
\item We implement an asynchronous computation pipeline and hierarchical beam-search retrieval, enabling efficient large-scale deployment across heterogeneous hardware.
\end{itemize}

\textbf{Significance.} Long user behavior sequences contain rich signals of interest evolution and dependency, yet industrial recommender systems often fail to fully leverage them due to computational and latency constraints~\cite{pan2024survey}. \textbf{MuSeR} unifies \textit{long-sequence compression}, \textit{multi-interest modeling}, and \textit{multimodal alignment} within a deployable retrieval framework, effectively mitigating issues like interest forgetting and limited representation capacity. By balancing retrieval efficiency and personalization accuracy, MuSeR marks a key advancement in \textbf{production-ready recommender systems with stronger adaptability and expressive power.}

%% file: Contents/METHODOLOGY.tex
\section{Method}
\label{sec:method}

We introduce the technical details of the proposed \textbf{MuSeR} framework in this section.  
We begin with an overview of the overall architecture, followed by detailed explanations of its key components, including hierarchical temporal compression, multi-query interest extraction, and multimodal semantic alignment.  
Finally, we describe the inference and optimization.

\subsection{Overview}
This section briefly introduces the overall pipeline of \textbf{MuSeR}, with the framework overview illustrated in Figure~\ref{fig:architecture}.  
Building upon the industrial MGS framework mentioned in \cite{zhou2019deep}, where the implementation details were not elaborated, MGS can be regarded as a NANN-style retrieval system based on HNSW, supporting multi-objective indexing and search. Based on this retrieval architecture, MuSeR further unifies ultra-long user behavior modeling, multi-interest representation learning, and multimodal alignment within a single framework.   
The system follows a hybrid offline–online computation strategy: offline modules asynchronously encode users’ long-term behavioral sequences into multiple interest embeddings, while the online module continuously updates short-term session behaviors for real-time retrieval.

Formally, given a user $u$ with an interaction sequence 
$S_u = [i_1, i_2, \dots, i_T]$ where $i_t \in \mathcal{I}$ denotes an interacted item,  
the learning objective is to estimate the matching function:
\begin{equation}
    \hat{y}_{ui} = F_\Theta(S_u, i),
\end{equation}
where $F_\Theta$ maps an ultra-long sequence into a user representation $\boldsymbol{u}$ and predicts the interaction probability between $u$ and item $i$.  
To achieve both accuracy and efficiency, MuSeR contains three major modeling phases described below.

\subsection{Hierarchical Temporal Compression}
\label{subsec:compress}

Ultra-long user histories (typically $10^4$–$10^5$ items) contain rich behavioral patterns but are expensive to process with Transformer-based models~\cite{vaswani2017attention}.  
We thus introduce a hierarchical temporal compression mechanism to preserve multi-scale temporal information while maintaining computational tractability.

\paragraph{Temporal segmentation.}
The full sequence $S_u$ is partitioned into three chronological segments:
\begin{equation}
S_u = [S_u^{(r)}, S_u^{(m)}, S_u^{(e)}],
\end{equation}
corresponding to recent ($S_u^{(r)}$), mid-term ($S_u^{(m)}$), and early ($S_u^{(e)}$) behaviors.  
The recent segment remains intact to maintain high-resolution dynamics, while earlier segments are gradually pooled with increasing stride to suppress redundancy.

\paragraph{Progressive pooling.}
We apply different down-sampling rates for different temporal regions:
\begin{align}
    S_u^{(r)} &= [i_1, \dots, i_{N_r}], \\
    S_u^{(m)} &= \mathrm{Pool}_{16}(i_{N_r+1:N_m}),  \\
    S_u^{(e)} &= \mathrm{Pool}_{64}(i_{N_m+1:T}),
\end{align}
where $\mathrm{Pool}_k(\cdot)$ denotes sum or attentive pooling over $k$ neighboring items.  
After compression, the total effective sequence length can be represented as:
\begin{equation}
L_c = N_r + \frac{N_m - N_r}{16} + \frac{T - N_m}{64}.
\end{equation}

\paragraph{Sequence encoding.}
Each token is represented as the sum of item and positional embeddings:
\begin{equation}
    \mathbf{x}_t = \mathbf{E}_{\text{id}}(i_t) + \mathbf{E}_{\text{pos}}(t),
\end{equation}
then encoded by a lightweight Transformer encoder:
\begin{equation}
    \mathbf{H}_u = \mathrm{TransformerEnc}([\mathbf{x}_1, \dots, \mathbf{x}_{L_c}]).
\end{equation}
The output \(\mathbf{H}_u \in \mathbb{R}^{L_c \times d}\) effectively captures both short-range and long-range dependencies, preserving long-term interest signals at low computational cost.

\subsection{Multi-Query Interest Extraction}
\label{subsec:multiquery}

User intent is often inherently multi-faceted: for example, the same user may simultaneously consume news, shop, or browse short videos. To uncover this heterogeneity, we introduce a multi-query attention module that decodes multiple interest embeddings from \(\mathbf{H}_u\).

A set of $M$ learnable query vectors $\{\boldsymbol{q}_1,\dots,\boldsymbol{q}_M\}$ attend to the contextual states through scaled dot-product attention:
\begin{equation}
\boldsymbol{u}_m = \mathrm{softmax}\!\left(\frac{\boldsymbol{q}_m \mathbf{H}_u^\top}{\sqrt{d}}\right)\!\mathbf{H}_u, 
\quad m = 1,\dots,M.
\end{equation}
Each $\boldsymbol{u}_m$ encodes a distinct behavioral aspect derived from different temporal or contextual patterns.  
During the offline training phase, all \(\{\boldsymbol{u}_m\}\) jointly predict future user behaviors; during online inference, the retrieval system adaptively selects the most relevant interest vector according to the given candidate item embedding \(\boldsymbol{e}_i\):
\begin{equation}
    \alpha_m = \frac{\exp(\mathrm{sim}(\boldsymbol{u}_m, \boldsymbol{e}_i)/\tau)}
          {\sum_n \exp(\mathrm{sim}(\boldsymbol{u}_n, \boldsymbol{e}_i)/\tau)}, \quad
    \boldsymbol{u} = \sum_m \alpha_m \boldsymbol{u}_m.
\end{equation}
Here, $\mathrm{sim}(\cdot)$ denotes cosine similarity and $\tau$ is a temperature parameter controlling interest sharpness.  

This mechanism allows MuSeR to align fine-grained interests with corresponding candidate categories dynamically.

\subsection{Multimodal Semantic Alignment}
\label{subsec:multimodal}

Purely ID-based representations lack semantic generalization and perform poorly for cold-start or sparse items.  
To mitigate this, we enhance items with multimodal semantic embeddings that align text- and content-level information with user representations.

Each item's textual description is first processed by a powerful large language model (ERNIE-4.0-Turbo) through chain-of-thought prompting, generating condensed and informative semantic perspectives.
We then distill this process into a lightweight variant (\texttt{ERNIE-Speed}) to scale cost-efficiently, and project the resulting summaries into dense semantic vectors using the BGE embedding model.
Finally, we fuse the content and ID embeddings as follows:
$
\boldsymbol{e}_i = \boldsymbol{W}_1\boldsymbol{e}_i^{(\text{id})}
+ \boldsymbol{W}_2\boldsymbol{e}_i^{(\text{sem})}.
$
This multimodal projection effectively maps user intent vectors and item semantics into a unified shared embedding space, facilitating semantically meaningful and accurate recalls across diverse recommendation scenarios.

\subsection{Training Objective}
\label{subsec:training}

MuSeR optimizes a next-$K$ behavior prediction objective, leveraging multiple interest embeddings to forecast users’ immediate future interactions.  
Given positive items $\mathcal{P}_u$ and sampled negatives $\mathcal{N}_u$, the training loss combines contrastive retrieval and orthogonality regularization as follows:
\begin{equation}
\scalebox{0.65}{$\displaystyle
\mathcal{L}_{\text{pred}}
= - \sum_{m=1}^{M} \sum_{i^+ \in \mathcal{P}_u}
   \log
   \frac{
      \exp\!\left(\mathrm{sim}(\boldsymbol{u}_m, \boldsymbol{e}_{i^+}) / \tau\right)
   }{
      \displaystyle
      \exp\!\left(\mathrm{sim}(\boldsymbol{u}_m, \boldsymbol{e}_{i^+}) / \tau\right)
      +
      \sum_{i^- \in \mathcal{N}_u}
      \exp\!\left(\mathrm{sim}(\boldsymbol{u}_m, \boldsymbol{e}_{i^-}) / \tau\right)
   }
$}
\label{eq:pred_loss}
\end{equation}
\begin{equation}
\mathcal{L}_{\text{orth}}
= \frac{1}{M^2}\!
   \sum_{m \ne n}
   \left\|
      \boldsymbol{u}_m^{\top}\boldsymbol{u}_n
   \right\|_2^2
\label{eq:orth_loss}
\end{equation}
\begin{equation}
\mathcal{L}
= \mathcal{L}_{\text{pred}}
 + \lambda\,\mathcal{L}_{\text{orth}}
\label{eq:total_loss}
\end{equation}
The first term maximizes mutual similarity between interest representations and ground-truth positives, while the second encourages decorrelation across different interest heads to ensure diversity.  
Training uses Adam with warm-up, dropout, and mixed precision to maintain convergence and efficiency.





%% file: Contents/Deployment.tex
\section{Online Deployment}
\label{sec:deployment}

\begin{figure}[t]
    \centering
    \includegraphics[width=0.95\linewidth]{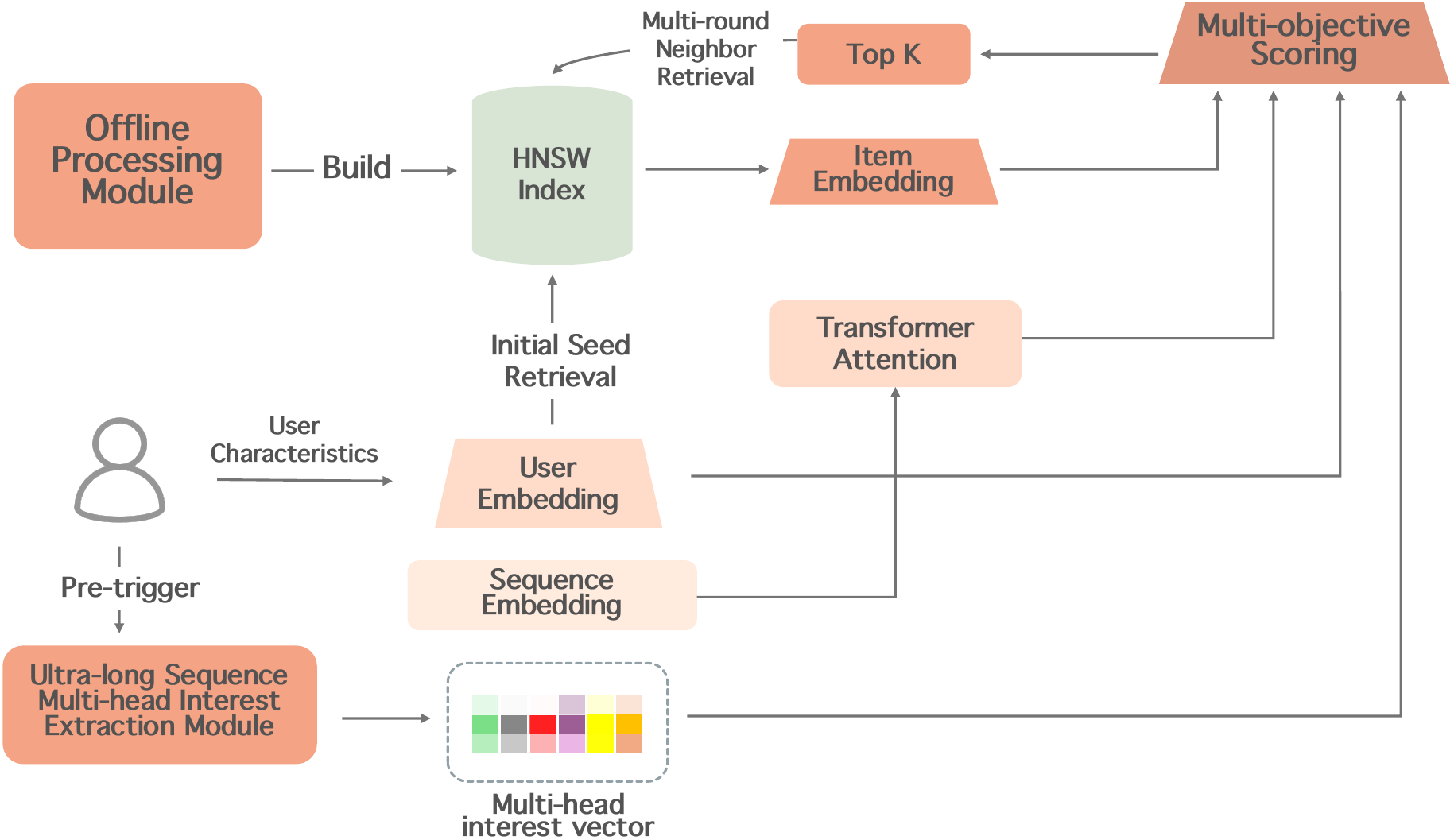}
    \caption{Overview of the online deployment architecture. The framework integrates asynchronous ultra-long sequence modeling with multi-objective retrieval. The offline processing module constructs the HNSW index, while real-time user embeddings and multi-head interest extraction interact with transformer attention and a multi-objective scoring network for efficient large-scale retrieval.}
    \label{fig:deployment}
\end{figure}

This section outlines the online deployment framework of MuSeR, highlighting the integration of \textbf{asynchronous computation} for ultra-long sequence modeling and \textbf{multi-objective retrieval} optimization in large-scale industrial environments. As illustrated in Fig.~\ref{fig:deployment}, the proposed architecture combines offline index construction, asynchronous user representation updates, and multi-objective scoring to achieve efficient and scalable online retrieval.

\subsection{Asynchronous Computation for Ultra-long Sequences}
In the industrial environment, user behavior sequences often span tens of thousands of interactions, making real-time modeling computationally prohibitive.  
To address this challenge, we deploy an \textbf{asynchronous cache-based computation framework}.  
The ultra-long sequence encoder and multi-head interest extraction module operate under a dual-frequency mechanism: high-frequency updates for recent short-term sequences and low-frequency asynchronous refresh for long-term interest embeddings.  

Specifically, user representations $\{\boldsymbol{u}_1^{(\text{long})}, \ldots, \boldsymbol{u}_M^{(\text{long})}\}$ are computed offline or refreshed periodically according to traffic load. These embeddings are stored in distributed caches and reused during online inference.  
Real-time requests only trigger the rapid computation of short-term sequence encoder outputs $\boldsymbol{u}^{(\text{short})}$, which are dynamically fused with cached embeddings via a weighted aggregator as follows:
\begin{equation}
\boldsymbol{u}^{(\text{final})} = \beta(t)\,\boldsymbol{u}^{(\text{short})} + [1-\beta(t)]\,\boldsymbol{u}^{(\text{long})},
\end{equation}
where the fusion weight $\beta(t)$ is determined by adaptive load monitoring and service concurrency.  
The system dynamically throttles the QPS (queries per second) of asynchronous updates to maintain an optimal trade-off between model effectiveness and computational cost.  
This heterogeneous-frequency computation pipeline reduces redundant encoding by up to 60\%, ensuring both low latency and stable retrieval quality under real-world traffic fluctuations.

\subsection{Multi-objective Retrieval and Efficient Vector Indexing}
The online retrieval service extends the standard MGS framework with a multi-objective search paradigm inspired by HNSW~\cite{malkov2018hnsw}.
We construct a multi-layer hierarchical vector index graph using Euclidean distance as the primary similarity criterion for candidate linkage and neighbor selection.
During retrieval, a complex DNN-based scoring network jointly evaluates multiple objectives such as semantic relevance, user engagement probability, and commercial conversion potential, enabling more precise and balanced candidate ranking.

A key challenge arises when the index-building metric (Euclidean distance) differs from the ranking metric of the multi-objective model, causing recall inefficiency.  
To mitigate this misalignment, we redesign the retrieval process as a \textbf{hierarchical beam-search strategy}.  
At each layer, candidate nodes are expanded by top-$k$ anchors and re-scored with fine-grained objectives.  
The beam width and expansion ratio are adaptively tuned through parameter exploration, aligning graph topology with model scoring.  
This improvement significantly reduces recall loss while maintaining high throughput.
Furthermore, the proposed retrieval mechanism efficiently exploits heterogeneous hardware environments.  
The graph index construction runs on high-capacity CPU clusters, while real-time vector traversal and scoring are executed on GPU or Kunlun NPUs with mixed-precision acceleration.  
The new hierarchical beam-search approach automatically balances computation across tiers, achieving hardware-level parallelism with minimal memory overhead.  
Empirically, this strategy effectively reduces end-to-end retrieval latency by 27\% and lowers overall engineering deployment cost by over 35\% compared with conventional flat HNSW indexing.

\subsection{Representation and Cross-model Fusion}
Our retrieval architecture combines the \textbf{dual-tower model} and \textbf{deep cross-network} advantages into one unified retrieval system.  
The document tower fully utilizes side information (e.g., category, semantic, and multimodal signals), producing compact embeddings for high-throughput retrieval.  
Meanwhile, the user tower, equipped with multi-head interest extraction, provides diverse high-resolution behavioral vectors.  
The DNN interaction layer within the multi-objective scoring model effectively captures cross-feature dependencies such as historical–semantic alignments.  
This hybrid design enhances modeling precision and extends the recall boundary, substantially increasing relevance coverage and improving retrieval gains in online production.

Overall, the deployed online system realizes an \textbf{asynchronous, hierarchical, and multi-objective retrieval framework} that maintains millisecond-level latency while sustaining billion-scale indexing.  
The combination of cache-based asynchronous updates and adaptive beam-search retrieval enables flexible trade-offs between recall quality and system cost, ensuring robust and efficient recommendation performance in large-scale industrial settings.

\subsection{Production Deployment Details}
\label{app:deploy}
In early 2025, the MuSeR framework was rolled out to the production environment
of the Baidu APP, first serving the homepage recommendation stream and then
being extended to the discovery feed and the short-video feed. This rollout was
validated through large-scale online A/B testing across these core traffic
scenarios. Experiments were conducted on live traffic with user-level bucketing,
where buckets are assigned by a hash of the user ID and held fixed throughout
the experiment window; control and treatment differ only in the retrieval
channel under study, while all downstream ranking and mixing stages remain
identical. Each experiment ran continuously for multiple weeks over tens of
millions of daily users per bucket, and the reported lifts are statistically
significant under a two-sided test. In the same experiment on the homepage feed,
MuSeR increased feed Daily Active Users (DAU) by \textbf{+0.26\%} and total
session duration by \textbf{+0.89\%}, indicating that the framework improves not
only immediate interaction outcomes but also sustained user engagement. At this
traffic scale, and given that the feed is already served by a mature ensemble of
retrieval sources, a gain of this magnitude from a single retrieval channel is
a meaningful improvement. These results confirm that MuSeR's unified
architecture, combining \textbf{hierarchical temporal compression} over
per-user histories of up to $10^{5}$ interactions with \textbf{multi-modal
semantic alignment}, can translate modeling advances into measurable business
impact under strict industrial constraints (e.g., latency budgets,
heterogeneous hardware, and high-QPS traffic).

Concretely, MuSeR improves retrieval quality by jointly addressing two
long-standing limitations in production recommenders. First, it mitigates
long-history underutilization by compressing ultra-long sequences in a
time-aware, information-preserving manner, enabling the system to exploit
stable long-term preferences without exceeding online compute budgets. In
production, long-term user representations are computed off the serving path
and cached, so the ultra-long history never enters the request-time compute
budget. Second, it reduces the semantic gap inherent in sparse ID-based
representations through multi-modal alignment, allowing the retriever to better
capture content similarity and user intent at a semantic level, which is
particularly useful for recently published items, rich-media content, and
cross-domain consumption patterns. Together, these capabilities enable MuSeR to
provide higher-recall candidate sets with stronger relevance and robustness,
which in turn contributes to the observed improvements in engagement metrics.

The specific application scenarios are illustrated below. Figure~\ref{fig:scene_home}
shows the homepage ``Recommended'' feed, which presents a mixture of long-form
news articles and rich-media cards. In this context, MuSeR leverages long-term
user histories to infer enduring interests (e.g., recurring topics, preferred
sources, and stable content styles) while still remaining responsive to recent
interactions. By combining cached long-term representations with fine-grained
short-term signals, MuSeR delivers personalized recommendations that retain
topical diversity, supporting broader exploration within a user's interest
space. Figure~\ref{fig:scene_short} depicts the discovery feed and short-video
grid, a highly dynamic environment characterized by rapid intent shifts, short
feedback cycles, and strong sensitivity to freshness and context. Here, MuSeR's
ability to fuse real-time session signals with cached long-term interests is
crucial for maintaining both instant relevance and recall stability as user
goals change quickly (e.g., from entertainment to information seeking). Moreover,
the use of multi-modal representations strengthens matching between users and
rich video candidates beyond sparse IDs, improving semantic recall under
heterogeneous modalities; this effect is more pronounced in the short-video
setting, where item lifecycles are short and per-item interaction histories are
correspondingly thin.

%% file: Contents/EXPERIMENT.tex
\input{Contents/Table}
\section{Experiments}
We present experimental settings and extensive empirical results of \textbf{MuSeR} in
this section. 
\subsection{Experiment Settings}
\noindent\textbf{Dataset.} We evaluate on three subsets of the Amazon 2023 review data (Instruments, Video Games, Industrial \& Scientific)\cite{mcauley2023amazon,huggingface_amazon2023}. We apply 5-core filtering\cite{he2017neuralmf}, build temporally ordered sequences truncated/padded to length 50, and split the data chronologically into 70\%/10\%/20\% for train/validation/test. We also report results on a large-scale proprietary dataset from Baidu's information feed, containing one month of real-world user--item interaction logs with about 100M users and 10M items; we train on the first 29 days and test on the last day.

\noindent\textbf{Model.} Our model is a 6-layer Transformer encoder (hidden size $d{=}512$, 8 attention heads, dropout 0.1)~\cite{vaswani2017attention} with 6-query interest extraction (temperature $\tau{=}0.07$), sinusoidal positional encodings (max length 10k)~\cite{vaswani2017attention}, and multimodal fusion over ID and content features. For text/content semantics, we distill ERNIE-4.0-Turbo summaries and use BGE-base-en embeddings (768d)~\cite{xiao2023bge} as well as Visualized-BGE/VISTA for hybrid multimodal alignment when applicable~\cite{zhou2024vista}. Fusion weights are initialized to 0.7/0.3 and learned end-to-end. We include an orthogonality regularization on multi-interest heads with coefficient $\lambda{=}0.01$ (a common technique to decorrelate factors in multi-head/mixture settings, also adopted in multi-interest retrieval literature such as \cite{liu2024kuaiformer}).

\noindent\textbf{Optimization and training.} We use AdamW~\cite{loshchilov2019adamw} with initial learning rate $1\mathrm{e}{-4}$, weight decay $1\mathrm{e}{-5}$, cosine-annealing schedule with 1k warm-up steps~\cite{loshchilov2017sgdr}, gradient clipping at 1.0~\cite{pascanu2013difficulty}, and batch sizes of 256 (public datasets) and 4096 (industrial datasets), employing mixed precision when available. Public baselines are tuned under a unified protocol.

\noindent\textbf{Baselines.} We compare with Caser~\cite{tang2018caser}, GRU4Rec~\cite{hidasi2016gru4rec}, SASRec~\cite{kang2018sasrec}, BERT4Rec~\cite{sun2019bert4rec}, FMLP-Rec~\cite{zhou2022fmlprec}, S$^3$-Rec~\cite{zhou2020s3rec}, TIGER~\cite{zhu2023tiger}, and NANN \cite{chen2022nann} using identical preprocessing and splits. Each method is run with 5 random seeds; we conduct two-sided $t$-tests with $p{<}0.05$ for statistical significance.

\noindent\textbf{Infrastructure.} Our implementation uses PyTorch 1.13.1~\cite{paszke2019pytorch}, HuggingFace Transformers~\cite{wolf-etal-2020-transformers} for text components, and Horovod for distributed training~\cite{sergeev2018horovod} with NCCL backend and CUDA/cuDNN. Mixed-precision training uses NVIDIA Apex when applicable~. Inference for the long-sequence encoder runs on NVIDIA A10 (18 cards) accelerated by TensorRT FP16. Training is conducted on NVIDIA A100 (4 cards) and Baidu Kunlun P800 NPUs; large-scale training scales to H20$\times$8 or Kunlun P800$\times$16. Retrieval employs Kunlun R200 (293{+}151) with PaddlePaddle/PaddleInference for serving. The asynchronous computation pipeline uses Redis for distributed caching and Apache Kafka for streaming, with Consul for service discovery.

\noindent\textbf{Metrics.} Beyond standard Recall@K and NDCG@K, we also report MRR, HR, and Coverage to assess ranking quality and catalog utilization. For industrial scenarios, we further track online metrics including CTR, CVR, and engagement signals such as dwell time and session duration. 

We also report results on a large Baidu APP feed dataset comprising $\sim$100M active users and 10M items, yielding $\sim$5B daily interactions. We use days 1–29 for training and day 30 for testing, and evaluate in two realistic scenarios: Home Feed (personalized long-term engagement) and Discovery/Short-Video (fast-shifting short-term intents). Online metrics include CTR/CVR and engagement (dwell time, session duration), in addition to offline retrieval metrics.


\subsection{Performance Analysis on Public Datasets}

As shown in Table~\ref{tab:baseline_model_performance}, the experimental results on three public benchmark datasets demonstrate the superior performance of MuSeR across diverse recommendation scenarios. On the \textbf{Instrument} dataset, MuSeR achieves the best performance across all metrics, with Recall@5 of 0.0381, Recall@10 of 0.0587, NDCG@5 of 0.0258, and NDCG@10 of 0.0319, showing statistically significant improvements over the strongest baseline NANN in most metrics ($p < 0.05$). 
Similarly, on the \textbf{Scientific} dataset, MuSeR consistently outperforms all competing methods, achieving Recall@5 of 0.0288, Recall@10 of 0.0441, NDCG@5 of 0.0188, and NDCG@10 of 0.0238, with statistically significant gains over NANN in Recall@5, Recall@10, and NDCG@5. For the \textbf{Game} dataset, MuSeR again establishes new state-of-the-art results with Recall@10 of 0.0919, NDCG@5 of 0.0389, and NDCG@10 of 0.0488, demonstrating significant improvements over the second-best performer, NANN.

The consistent improvements across all three datasets validate several key aspects of MuSeR's design. First, the hierarchical temporal compression mechanism effectively captures both short-term and long-term user preferences without losing critical behavioral signals. Second, the multi-query interest extraction successfully disentangles diverse user intents, enabling more accurate representation of complex user behavior patterns. Third, the multimodal semantic alignment enhances the model's ability to understand item semantics beyond sparse ID embeddings, particularly beneficial for cold-start scenarios and cross-domain recommendations.
Notably, MuSeR's performance gains are most pronounced on metrics like Recall@5 and NDCG@5, indicating that the framework excels at identifying the most relevant items for users, a crucial capability for practical recommendation systems where top-ranked results have the highest impact on user satisfaction. The statistical significance of these consistent improvements across multiple benchmark datasets confirms that MuSeR's unified approach to long-sequence modeling, multi-interest representation, and multimodal alignment provides robust and broadly generalizable benefits for sequential recommendation.

\subsection{Overall Performance on the Private Dataset}
On Baidu’s large-scale private industrial dataset, \textbf{MuSeR} consistently achieves the strongest recall across both the Home Feed and the Discovery and Short Video scenarios. Compared with established baselines including SASRec, NANN, TIGER, and KuaiFormer, MuSeR delivers higher Recall@100 and Recall@500 in both settings, as reported in Table~\ref{tab:feed_overall} and Table~\ref{tab:short_overall}. These results substantiate that unifying hierarchical temporal compression, multi-query interest extraction, and multimodal semantic alignment can simultaneously preserve long-horizon behavioral coverage and high-resolution short-term intent modeling. This unified design expands the effective recall boundary under industrial latency and cost constraints by reducing redundancy in ultra-long sequences while retaining diverse intent signals. In the Discovery and Short Video scenario, where user preferences evolve rapidly and items span heterogeneous modalities, the gains are pronounced, indicating that MuSeR better disentangles conflicting interests and aligns sparse identifiers with semantic content.

Overall, the consistent improvements in Table~\ref{tab:feed_overall} and Table~\ref{tab:short_overall} affirm that scalable long-sequence modeling with multi-interest representations is crucial for production recommendation, as it mitigates interest forgetting, strengthens cold-start robustness, and provides a higher-quality candidate pool for downstream ranking without sacrificing efficiency. As shown by these Tables, MuSeR establishes a new state of the art on both Recall@100 and Recall@500 across scenarios. These improvements confirm that capturing multi-scale temporal patterns and disentangling user intents are essential for robust recall in real-world traffic, while multimodal semantic alignment enhances coverage for sparse and newly published items. The resulting candidate sets enable downstream re-rankers to operate on more relevant pools, ultimately contributing to better end-to-end personalization quality under real-world production constraints.

\begin{figure*}[htbp]
    \centering
    \begin{subfigure}[b]{0.32\textwidth}
        \centering
        \includegraphics[width=\textwidth]{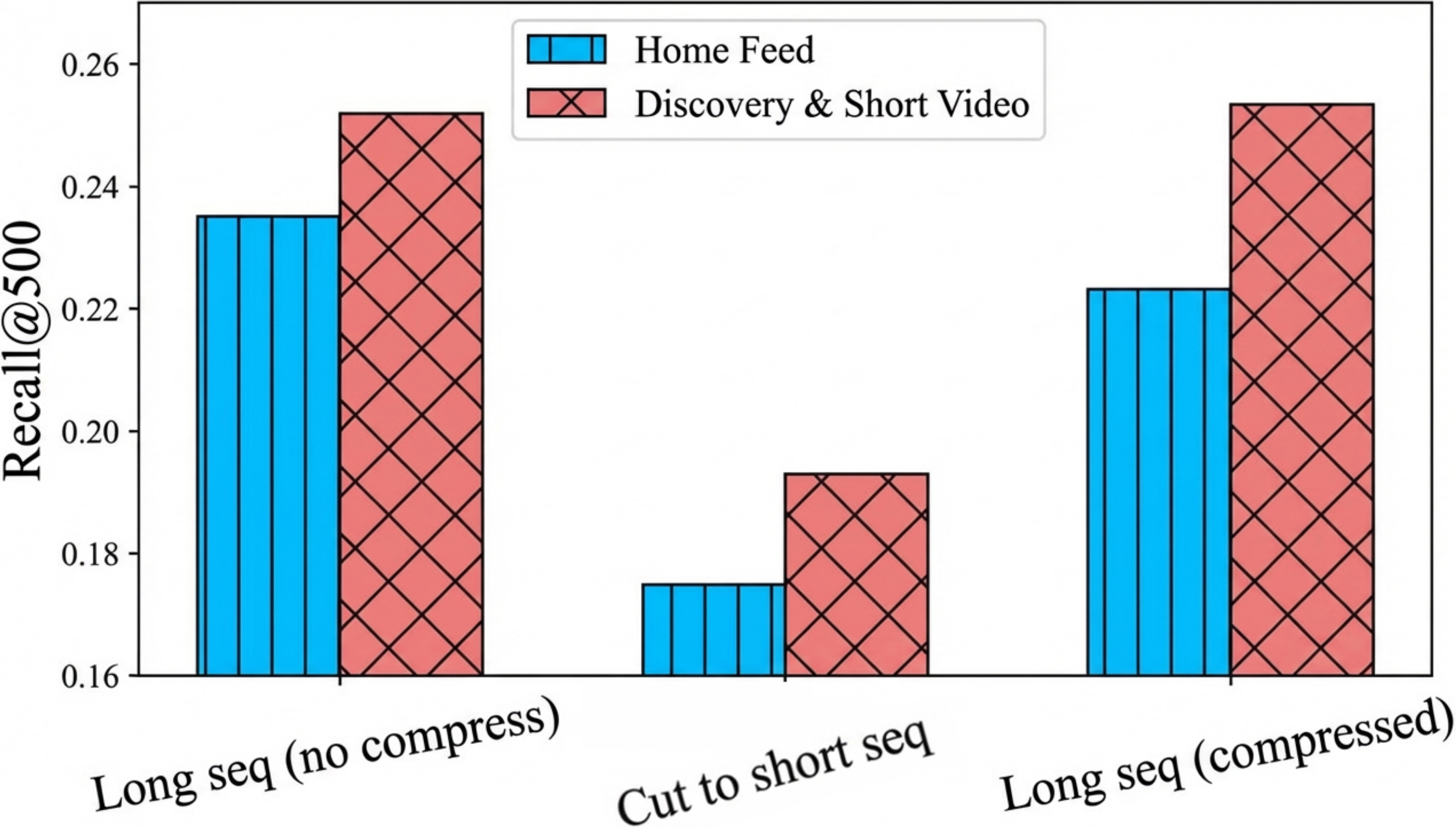}
        \caption{Sequence processing strategies}
        \label{fig:ablation_sequence}
    \end{subfigure}
    \hfill
    \begin{subfigure}[b]{0.32\textwidth}
        \centering
        \includegraphics[width=\textwidth]{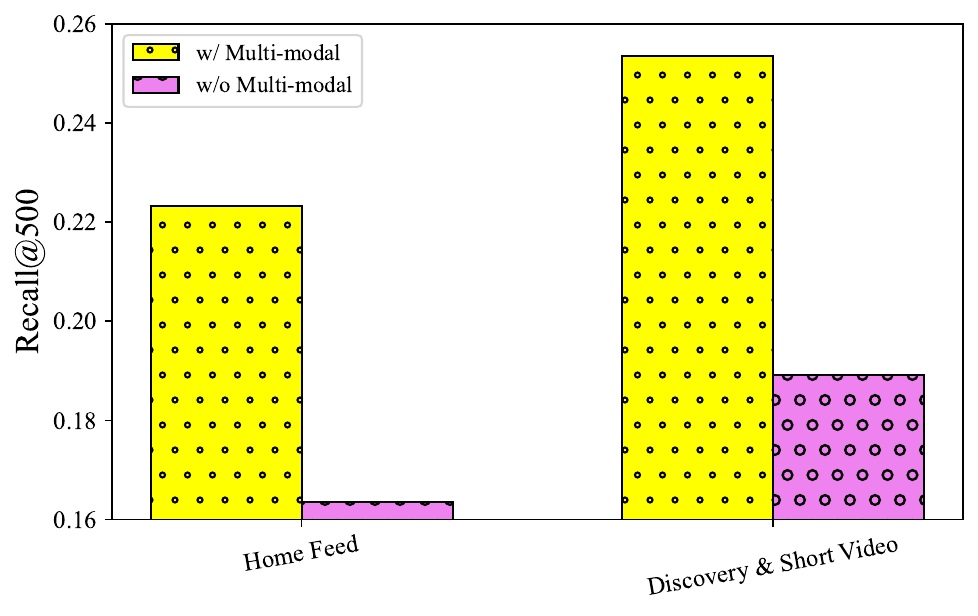}
        \caption{Multimodal components}
        \label{fig:ablation_modality}
    \end{subfigure}
    \hfill
    \begin{subfigure}[b]{0.32\textwidth}
        \centering
        \includegraphics[width=\textwidth]{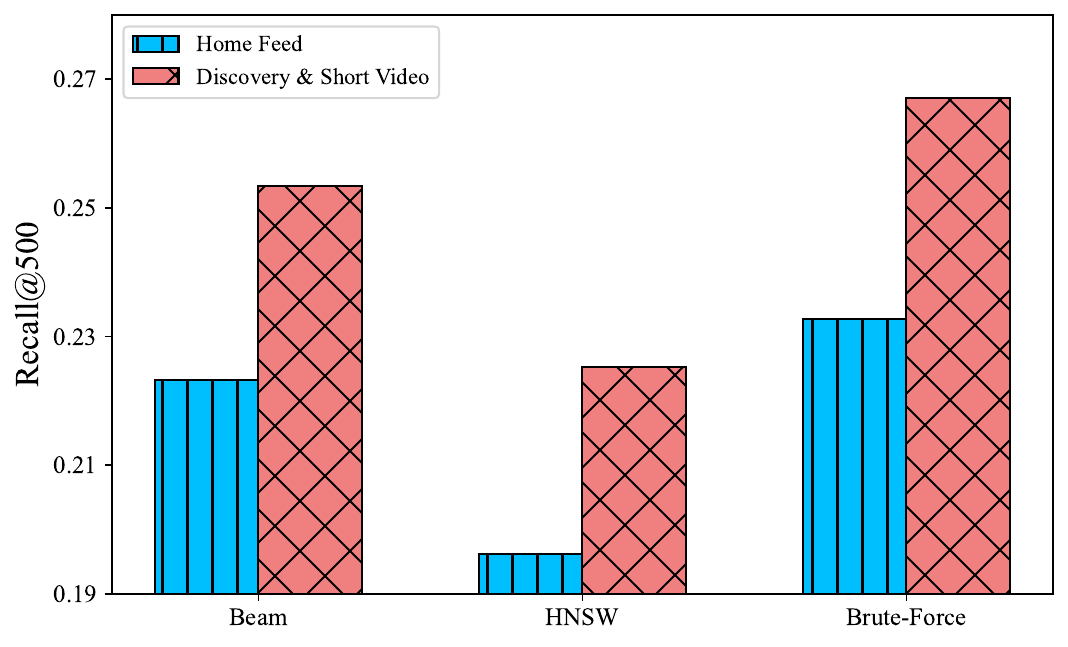}
        \caption{Retrieval strategies}
        \label{fig:retrieval}
    \end{subfigure}
    \caption{Recsys study results.}
    \label{fig:ablation_study}
\end{figure*}

\begin{figure*}[t!]
    \centering
    \begin{subfigure}[b]{0.45\textwidth}
        \centering
        \includegraphics[width=\textwidth]{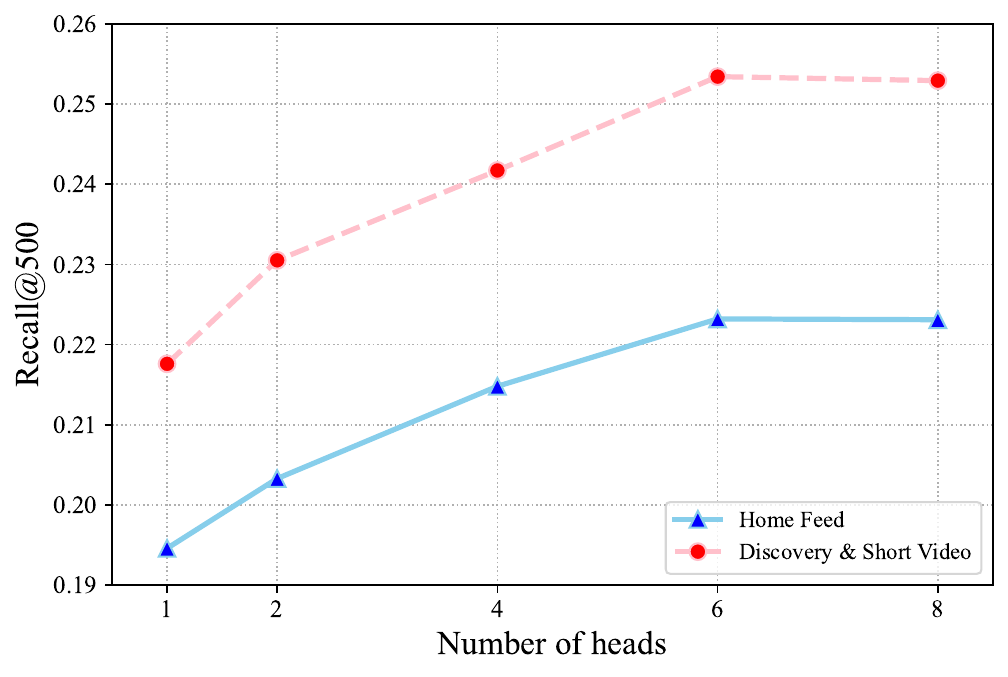}
        \caption{Number of interest heads}
        \label{fig:interest_heads}
    \end{subfigure}
    \hfill
    \begin{subfigure}[b]{0.45\textwidth}
        \centering
        \includegraphics[width=\textwidth]{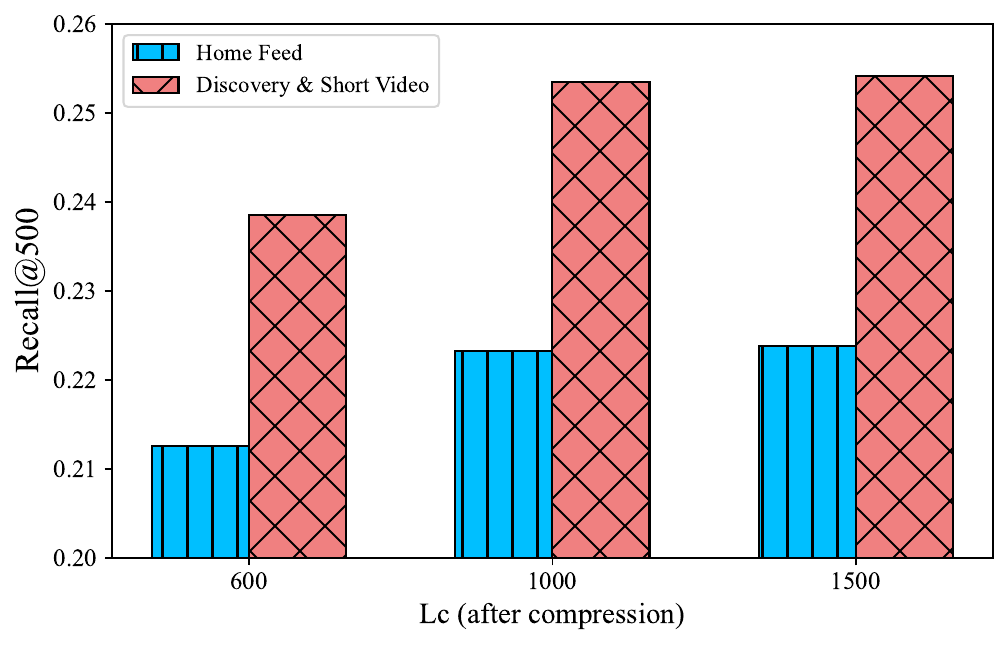}
        \caption{Compression length (Lc)}
        \label{fig:lc_length}
    \end{subfigure}
    \caption{Hyperparameter sensitivity analysis. (a) The optimal number of interest heads is around 4-6 for both scenarios, balancing model capacity and overfitting risks. (b) Compression length analysis indicates that Lc=1000 provides the best trade-off between information preservation and computational efficiency.}
    \label{fig:hyperparameter_analysis}
\end{figure*}

\begin{table}[t]
\centering
\caption{Quantitative comparison on the Home Feed scenario. Best results are in bold.}
\label{tab:feed_overall}
\begin{tabular}{lcc}
\toprule
Method & Recall@100 & Recall@500 \\
\midrule
SASRec~\cite{kang2018sasrec}          & 0.0389 & 0.1176 \\
NANN~\cite{chen2022nann}     & 0.0714 & 0.1982 \\
TIGER~\cite{rajput2023tiger}  & 0.0511 & 0.1654 \\
KuaiFormer                          & 0.0524 & 0.1701 \\
\textbf{Ours (MuSeR)}               & \textbf{0.0870} & \textbf{0.2232} \\
\bottomrule
\end{tabular}
\end{table}

\begin{table}[t]
\centering
\caption{Quantitative comparison on the Discovery and Short Video scenario. Best results are in bold.}
\label{tab:short_overall}
\begin{tabular}{lcc}
\toprule
Method & Recall@100 & Recall@500 \\
\midrule
SASRec~\cite{kang2018sasrec}          & 0.0569 & 0.1372 \\
NANN~\cite{chen2022nann}     & 0.1012 & 0.2287 \\
TIGER~\cite{rajput2023tiger}  & 0.0721 & 0.1881 \\
KuaiFormer                          & 0.0818 & 0.1932 \\
\textbf{Ours (MuSeR)}               & \textbf{0.1377} & \textbf{0.2534} \\
\bottomrule
\end{tabular}
\vspace{1mm}
\end{table}

\subsection{Ablation Study}

To understand the contribution of each key component, we first examine different sequence processing strategies. As shown in Figure~\ref{fig:ablation_sequence}, comparing long sequence without compression (0.2351), truncation to short sequence (0.1749), and our proposed hierarchical compression (0.2232) reveals several important insights. The significant performance drop with truncation (-0.0602) demonstrates the critical importance of preserving long-term behavioral signals. While compression introduces a slight performance reduction compared to uncompressed sequences (-0.0119), it enables practical deployment with an acceptable accuracy-efficiency trade-off. This finding validates our design choice of hierarchical compression as a balanced solution for industrial-scale deployment.

The impact of multimodal semantic alignment proves particularly striking in our ablation experiments. Figure~\ref{fig:ablation_modality} demonstrates substantial improvements when incorporating multimodal components, with Home Feed showing a 36.4\% relative improvement (from 0.1636 to 0.2232) and Discovery demonstrating a 33.9\% enhancement (from 0.1892 to 0.2534). These remarkable gains confirm our hypothesis that enriching sparse ID embeddings with semantic signals significantly enhances recommendation quality. The improvement is particularly pronounced in cross-modal scenarios like Discovery, where content understanding plays a crucial role in user preference modeling.
In our evaluation of retrieval strategies, we compare three distinct approaches: Brute-Force search, Beam search, and HNSW. As illustrated in Figure~\ref{fig:retrieval}, while Brute-Force achieves the highest recall scores (Home: 0.2328, Discovery: 0.2671), it comes with prohibitive computational costs for industrial deployment. Beam search emerges as a particularly compelling solution, maintaining strong performance (Home: 0.2232, Discovery: 0.2534). HNSW offers the fastest retrieval but with a moderate sacrifice in overall recall quality.

\subsection{Hyperparameter Analysis}
\label{sec:hyper}
The configuration of sequence length plays a crucial role in balancing model effectiveness and computational efficiency. We examine the impact of compressed sequence length by varying $N_r$ (recent sequence) and $N_m$ (mid-term sequence) while fixing $\text{Pool}_k$ parameters ($k_1=16$, $k_2=64$). As shown in Figure~\ref{fig:lc_length}, increasing the compressed length from 600 to 1000 yields clear gains, with Home Feed improving from 0.2126 to 0.2232 and Discovery improving from 0.2385 to 0.2534. This indicates that moderately longer contexts still provide additional useful signals, especially for capturing recurring preferences and cross-session transitions. However, further extending the length to 1500 shows diminishing returns, with only 0.0006\textasciitilde{}0.0007 improvement, suggesting that the model reaches a performance plateau under the current compression granularity and serving constraints. Overall, the chosen configuration ($N_r=800$, $N_m=2000$) achieves robust performance across scenarios, reflecting a practical ``sweet spot'' where longer histories improve effectiveness while keeping inference cost and latency manageable.

We further study the multi-interest head configuration. Figure~\ref{fig:interest_heads} shows that multi-head mechanisms consistently outperform single-head baselines across all scenarios, confirming the importance of disentangling heterogeneous user intents. The best overall setting is 6 attention heads, achieving recall scores of 0.2232 for Home Feed and 0.2534 for Discovery. The Discovery scenario benefits more from multi-interest modeling than Home Feed, showing a larger relative improvement (0.0358 vs.\ 0.0286), which aligns with its higher content diversity and faster intent shifts. When the number of heads exceeds 6, performance plateaus or slightly decreases, implying that overly fine-grained interest decomposition can increase optimization difficulty and introduce redundancy. In practice, this suggests that a moderate number of heads provides the best trade-off between representation capacity and model complexity for stable online deployment.

%% file: Contents/Table.tex
\begin{table*}[t]
\centering
\caption{Performance comparison of different models on three public datasets, including Instrument, Scientific, and Game. The best and second-best results are highlighted in bold and underlined font, respectively. ``*'' denotes statistically significant improvements (i.e., two-sided \textit{t}-test with $p < 0.05$) over the best baseline.}
\label{tab:baseline_model_performance}
\resizebox{\textwidth}{!}{
\begin{tabular}{l|cccc|cccc|cccc}
\toprule
\textbf{Methods} &
\multicolumn{4}{c|}{\textbf{Instrument}} &
\multicolumn{4}{c|}{\textbf{Scientific}} &
\multicolumn{4}{c}{\textbf{Game}} \\
\cmidrule(lr){2-5} \cmidrule(lr){6-9} \cmidrule(lr){10-13}
 & \textbf{Recall@5} & \textbf{Recall@10} & \textbf{NDCG@5} & \textbf{NDCG@10} 
 & \textbf{Recall@5} & \textbf{Recall@10} & \textbf{NDCG@5} & \textbf{NDCG@10} 
 & \textbf{Recall@5} & \textbf{Recall@10} & \textbf{NDCG@5} & \textbf{NDCG@10} \\
\midrule
Caser & 0.0242 & 0.0392 & 0.0154 & 0.0202 & 0.0172 & 0.0281 & 0.0107 & 0.0142 & 0.0346 & 0.0567 & 0.0221 & 0.0291 \\
GRU4Rec & 0.0345 & 0.0537 & 0.0220 & 0.0281 & 0.0221 & 0.0353 & 0.0144 & 0.0186 & 0.0522 & 0.0831 & 0.0337 & 0.0436 \\
HGN & 0.0319 & 0.0515 & 0.0202 & 0.0265 & 0.0220 & 0.0356 & 0.0138 & 0.0182 & 0.0423 & 0.0694 & 0.0266 & 0.0353 \\
SASRec & 0.0341 & 0.0530 & 0.0217 & 0.0277 & 0.0256 & 0.0406 & 0.0147 & 0.0195 & 0.0517 & 0.0821 & 0.0329 & 0.0426 \\
BERT4Rec & 0.0305 & 0.0483 & 0.0196 & 0.0253 & 0.0180 & 0.0300 & 0.0113 & 0.0151 & 0.0453 & 0.0716 & 0.0294 & 0.0378 \\
FMLP4-Rec & 0.0328 & 0.0529 & 0.0206 & 0.0271 & 0.0248 & 0.0388 & 0.0158 & 0.0203 & 0.0535 & 0.0860 & 0.0331 & 0.0435 \\
FDSA & 0.0364 & 0.0557 & 0.0233 & 0.0295 & 0.0261 & 0.0391 & 0.0174 & 0.0216 & 0.0485 & 0.0857 & 0.0353 & 0.0453 \\
S$^3$-Rec & 0.0340 & 0.0538 & 0.0218 & 0.0282 & 0.0253 & 0.0410 & 0.0172 & 0.0218 & 0.0533 & 0.0823 & 0.0351 & 0.0444 \\
\midrule
SID & 0.0319 & 0.0438 & 0.0237 & 0.0275 & 0.0155 & 0.0234 & 0.0103 & 0.0129 & 0.0480 & 0.0693 & 0.0333 & 0.0401 \\
CID & 0.0352 & 0.0507 & 0.0234 & 0.0285 & 0.0192 & 0.0300 & 0.0123 & 0.0155 & 0.0497 & 0.0748 & 0.0343 & 0.0424 \\
TIGER & 0.0368 & 0.0574 & 0.0242 & 0.0308 & 0.0275 & 0.0431 & 0.0181 & 0.0231 & 0.0570 & 0.0895 & 0.0370 & 0.0471 \\
TIGER-SAS & 0.0375 & 0.0576 & 0.0242 & 0.0306 & 0.0272 & 0.0435 & 0.0174 & 0.0227 & 0.0561 & 0.0891 & 0.0363 & 0.0469 \\
LETTER & 0.0372 & 0.0581 & 0.0243 & 0.0310 & 0.0276 & 0.0433 & 0.0179 & 0.0230 & 0.0576 & 0.0901 & 0.0373 & 0.0475 \\
NANN & \underline{0.0373} & \underline{0.0584} & \underline{0.0250} & \underline{0.0313} & \underline{0.0281} & \underline{0.0435} & \underline{0.0185} & \underline{0.0235} & \underline{0.0582} & \underline{0.0910} & \underline{0.0381} & \underline{0.0481} \\
\textbf{MuSeR} & \textbf{0.0381}$^*$ & \textbf{0.0587} & \textbf{0.0258}$^*$ & \textbf{0.0319}$^*$ & \textbf{0.0288}$^*$ & \textbf{0.0441}$^*$ & \textbf{0.0188}$^*$ & \textbf{0.0238} & \textbf{0.0592} & \textbf{0.0919}$^*$ & \textbf{0.0389}$^*$ & \textbf{0.0488}$^*$ \\
\bottomrule
\end{tabular}}
\end{table*}

%% file: Contents/RELATEDWORKS.tex
\section{Related Works}
\subsection{Sequence Modeling}
Recent work tackles long-sequence recommendation through efficiency-oriented pipelines and scalable architectures. Two-stage designs such as SIM~\cite{pi2020sim} and TWIN~\cite{chang2023twin} improve serving efficiency by first retrieving a target-relevant sub-sequence and then applying expressive attention over the filtered history; because the retained history is ultimately summarized into a single user representation, however, concurrent and conflicting interests remain only weakly separated. LONGER~\cite{chai2025longer} scales Transformer-style modeling to substantially longer interaction sequences, yet heterogeneous interests and the gap between ID and content semantics are still difficult to reconcile under online latency constraints. Classical sequential models including GRU4Rec~\cite{hidasi2016gru4rec}, SASRec~\cite{kang2018sasrec}, and BERT4Rec~\cite{sun2019bert4rec} target comparatively short sequences, and their purely ID-based representations offer limited leverage for long-range dependencies and for items or users whose interaction history is still sparse.

Multi-interest modeling has consequently become a major research direction, from MIND~\cite{li2019multi}, which applies dynamic routing to extract multiple interest capsules from a behavior sequence, to controllable frameworks~\cite{cen2020controllable} that expose the number and granularity of extracted interest vectors as tunable factors. Subsequent efforts adopt contrastive objectives~\cite{zhang2022re4} to sharpen the boundaries between distinct interests, and target-aware distillation~\cite{wang2022target} to reconcile multi-interest training with single-vector inference at serving time. Related lines pursue disentanglement~\cite{du2024disentangled} to reduce redundancy across learned interests, capsule-based extraction~\cite{cheng2024accurate,tang2023towards} to capture compositional preferences, and multi-granularity modeling~\cite{tian2022when,meng2023coarse} to express intent at both coarse and fine levels, with query-aware methods forming a closely related thread~\cite{guo2023query,white2010predicting}. High-capacity architectures such as KuaiFormer~\cite{liu2024kuaiformer} and HSTU~\cite{zhai2024hstu} deliver strong accuracy but incur inference cost that is difficult to absorb in real-time retrieval. On the alignment side, VISTA~\cite{zhou2024vista} and BGE-M3~\cite{chen2024bgem3} yield stronger semantic representations, yet they are not designed for sequential recommendation and provide no mechanism for tracking how interests evolve over time. MuSeR instead couples hierarchical compression, multi-interest extraction, and semantic alignment in a single end-to-end framework, whereas prior work typically optimizes only one of these axes in isolation.

\subsection{System Optimization Techniques}
Industrial deployment must balance model accuracy against serving efficiency under stringent latency and throughput requirements. MARM~\cite{lv2024marm} examines memory--compute trade-offs for caching intermediate representations under strict SLAs, but assumes a single cached state per user; multi-interest serving instead maintains several embeddings per user, whose refresh cadence should arguably differ according to how quickly each interest drifts. Asynchronous refresh in practice must also absorb traffic volatility and hardware heterogeneity to avoid latency spikes at peak load, a requirement that receives comparatively little explicit treatment. On the retrieval side, HNSW~\cite{malkov2018hnsw} supports fast approximate nearest-neighbor search with sub-linear query complexity, yet its fixed similarity metric can diverge from production ranking criteria, particularly when multimodal document towers and multi-head user interests must be traded off jointly against relevance, diversity, and freshness. MuSeR addresses this through hierarchical beam search that re-scores candidate neighborhoods with learned multi-objective functions, and through asynchronous long- and short-term fusion with cache-aware refresh scheduling that adapts update intervals to monitored traffic, sustaining throughput under fluctuating QPS while preserving expressiveness within production latency budgets.

%% file: Contents/CONCLUSION.tex
\section{Conclusion}
We presented MuSeR, a scalable long-sequence recommendation framework that
unifies hierarchical temporal compression, multi-query interest extraction, and
multimodal semantic alignment in a production-ready retrieval pipeline. By
preserving per-user histories of up to $10^{5}$ interactions at multiple
temporal resolutions, disentangling heterogeneous intents into distinct interest
vectors, and enriching sparse IDs with textual semantics, MuSeR improves recall
coverage, robustness to intent drift, and recommendation quality for recently
published items. Experiments on public benchmarks and a large-scale Baidu
dataset, together with online A/B tests, show statistically significant gains in
feed DAU and total session duration while meeting production latency budgets.
Future work will explore adaptive compression, dynamic head allocation, and
retrieval--ranking co-design at billion-scale.

\section*{Acknowledgement}
This research was deployed in real-world business scenarios with the support of Baidu's Feed Recall team and Recommendation Engine team. This research was supported by the Hong Kong Institute of AI for Science (HKAI-Sci), City University of Hong Kong. This research was also partially supported by the National Natural Science Foundation of China (No. 62502404), the Hong Kong Research Grants Council (Research Impact Fund No. R1015-23, Collaborative Research Fund No. C1043-24GF, RGC Research Fellow Scheme No. RFS2627-1S03, General Research Fund No. 11218325 and No. 11212926), and the Institute of Digital Medicine of City University of Hong Kong (No. 9229503).